\documentclass[12pt]{article}
\title{CONCEPTS\, FOR\, A\, THEORY\, OF\, THE\,
ELECTROMAGNETIC\, FIELD\footnote{Lecture given at
the Meeting on Electromagnetism organized by the
Fondation Louis de Broglie in Peyresq,  September
2002.  }\\[1cm]}

\author{Bartolom\'e COLL\\[0.5cm]
 {\small Syst\`emes de r\'ef\'erence relativistes}\\
{\small SYRTE-CNRS Observatoire de Paris,}\\
{\small 61, avenue de l'Observatoire, F-75014 Paris,}\\
{\small mailto:bartolome.coll@obspm.fr}}

\date{}

\begin{document}

\maketitle

\vskip 1cm

%
%

\begin{abstract}

The object of this contribution is twofold. On one hand, it
rises some general questions concerning the definition of the
electromagnetic field and its intrinsic properties, and it
proposes concepts and ways to answer them. On the other hand,
and as an illustration of this analysis, a set of quadratic
equations for the electromagnetic field is presented, richer in
pure radiation solutions than the usual Maxwell equations, and
showing a striking property relating geometrical optics to all
the other Maxwell solutions.

\end{abstract}

P.A.C.S.: 03.50.De; 04.20.-q
\vspace{2mm}

%
%

\section{Introduction}
\vspace{2mm}

I think that the principal mission of the scientific culture is
not to increase our {\em knowledge,} as it is frequently stated,
but to ameliorate our {\em understanding} of the world. It is
clear that, although frequently connected, and sometimes
intimately, these two goals present deep differences in content
and in extent.

Scientific culture, like any other human culture, when absorbed
without reflection, is also alienation. In particular, the
present state of the classical electromagnetic theory shows
abundantly this feature. And it is trying to escape a little to
this alienation on the subject that my collaborators and myself
built up some concepts and results, of which a part is presented
here \cite{Collaborations}.

This contribution wishes to ameliorate our understanding of the
electromagnetic field, and consists of very simple concepts,
arguments and propositions. But, even simple, these elements
present a certain interest: they rise pertinent questions on
classical fields, outline answers, propose new formalisms and
lead directly to new equations with striking properties. It is
about these points that I would like to talk here.

Here we are interested by the {\em texture} of the {\em
physical} electromagnetic fields \cite{texture}.

The first problem that a field theory faces, be it
electromagnetic or not, is the adequacy between the physical
phenomena trying to be described by the theory and the solutions
that the theory provides. Some aspects of this problem, and in
particular the elements that make today not one-to-one the
relation between physical phenomena and field theories, are
commented in Section 2.

The second problem is that of the adequacy between the physical
quantity describing the field itself (here the electromagnetic
field) and the mathematical object chosen by the theory to
represent this quantity (vectors for the electric and magnetic
fields, anti-symmetric tensor for the total electromagnetic
one). Section 3 analyses briefly this problem and points out its
experimental character.

By texture of an electromagnetic field we mean the particular
link among the ingredients of the mathematical object chosen to
represent this electromagnetic field, as well as among its
gradients, specially the invariant or intrinsic ingredients.
After remembering the notions of {\em regular} and {\em pure
radiation} electromagnetic fields, those of {\em proper} energy
density of an electromagnetic field and of observer {\em at
rest} with respect to a regular electromagnetic field are
presented in Section 4. It also presents the intrinsic relation
between the electromagnetic field and its {\em principal}
directions, the {\em physical} interpretation of them in terms
of the electric and magnetic fields with respect to any
observer, and the one-to-one {\em decomposition} of the
electromagnetic energy tensor in two pure radiation components.
It follows a method allowing, for the first time, to obtain the
necessary and sufficient conditions to be imposed on an
electromagnetic field in order to admit eigen-directions with
prescribed differential properties. Some comments about the
general form of the {\em charge scaling} law and the {\em
superposition} law for any electromagnetic theory are added.

It is usually ignored that, whatever be the tensorial object
describing an ordinary electromagnetic field, a strong one {\em
could} change not only its {\em intensity} but even its {\em
specific} tensor character. In Section 5 a formal example is
given in which strong electromagnetic fields would be given by
Lorentz tensors, and some results necessary to handle this
eventuality are presented.

Finally, in Section 6, as a precise application of the above
concepts, a quadratic generalization of Maxwell equations is
presented. These new equations for the electromagnetic field
admit ``more light" (pure radiation solutions) than the Maxwell
ones, but {\em exactly} the same charged electromagnetic fields
(regular solutions). They have the striking property, on one
hand, of being a geometric optics approximation of Maxwell
equations and, on the other hand, of containing  the exact
Maxwell equations themselves.

\vspace{2mm}

%
%

\section{Phenomena and Field Theories}
\vspace{2mm}

\ \indent {\bf a)} For Newton ({\em resp.} Maxwell)
classical field theory, the field outside the sources in a
finite region of the space ({\em resp.} space-time) determines
completely the masses ({\em resp.} charges) that produce it, and
consequently the field everywhere ({\em resp.} in all the causal
past of the region). For this reason, we call here for short
{\em pure field theories} the local versions of these theories,
and of their possible local generalizations (including General
Relativity), at the exterior of the sources; that is to say, the
set of regular solutions in a local domain of the corresponding
differential equations in vacuum \cite{Hehl}
\cite{commentmatter}.

Denote by  $\,{\bf F}\,$  the set of  all physical phenomena of
a certain class and by  $\,{\bf S}\,$  the set of (physical or
unphysical) phenomena described by solutions to a field theory
for this class. Obviously, the set of physical phenomena
described by such a field theory, $\,{\bf F}\cap {\bf S}\,,$
generically differs from the whole set $\,{\bf F} \cup {\bf
S}\,$ : $\,{\bf F} \cap {\bf S}\,$ $\subset$ $\,{\bf F}\cup {\bf
S}\,.$   Thus, generically $\,{\bf F}\,$ contains a subset
$\,{\bf \widehat F}\,$ of physical phenomena not described by
the field theory, $\,{\bf \widehat F}\,$ $\equiv$ $\,{\bf F} -
{\bf F}\cap {\bf S}\,,$  as well as $\,{\bf S}\,$ contains a
subset $\,{\bf \widehat S}\,$  of solutions to the field theory
which are unphysical, $\,{\bf \widehat S}\,$ $\equiv$ $\,{\bf S}
- {\bf F}\cap {\bf S}\,,$ so that we have $\,{\bf F} \cup {\bf
S}\,$ $\equiv$ $\,{\bf \widehat F}\cup ({\bf F}\cap {\bf S})\cup
{\bf \widehat S}\,.$

Theories for which the subset  $\,{\bf \widehat F}\,$ vanishes
are called  {\em complete}.  Theories for which the subset
$\,{\bf \widehat S}\,$  vanishes are called  {\em strict}.
\vspace{4mm}

     {\bf b)} For non-complete theories, the analysis of the
subset  $\,{\bf \widehat F}\,$ of phenomena not described by
field theories sets out, in general, a problem at once of
observation, experimentation and inferential logic. But, in
particular, some cases may be analysed (at least partially) with
the standard methods of theoretical physics.
 Among these cases, one may mention the possible existence of
space-time metrics insensitive to the polarization of
gravitational waves. If one accepts that the curvature of the
space-time due to the presence of a gravitational field follows
a ``universal law of gravitational deformation", then one can
locally separate unambiguously the metric from the gravitational
field, which turns out to be given by a two-form
\cite{UniversalLaw}. One can then show that the polarization of
a gravitational wave is not ``detected" by the space-time
metric.

Another case able to be analysed theoretically is that of
charge-independent and non-radiation electromagnetic fields. One
knows that, under Maxwell equations, to shake a charged particle
allows to detach parts of its field, which get away at the
velocity of light (radiation field component of charged
accelerated particles). The hypothesis that strong
electromagnetic fields are better represented by Lorentz tensors
than by two-forms, offers the possibility to detach from
particles non-escaping (non-radiation) parts of the field.

Also, there exist arguments to suspect that the solutions to
Maxwell equations representing light could be insufficient. This
possible lack of sufficiency of Maxwellian light may be defined
precisely, and equations avoiding it may be obtained.

At present, of these three examples, the second one is the more
speculative \cite{disconnected},  and although in Section 5 we
give some elements to handle Lorentz tensors related to
two-forms, it will not be study in depth. The first one has been
briefly presented in \cite{UniversalLaw}, and needs a critical
and comparative analysis with the standard points of view.
Finally, the third example has been analised in
\cite{NonLiMaxEqs}, and  will be explained in some detail in the
last section of this paper.
\vspace{4mm}

    {\bf c)}  Contrarily to an extended opinion, the unphysical
solutions  $\,{\bf \widehat S}\,$ of non-strict field theories
fill in, roughly spoken, almost all the space of solutions of
the field  equations, the solutions $\,{\bf F}\cap {\bf S}\,,$
that describe physical phenomena being a set of ``very null"
measure \cite{verynull}.

    This situation is due to the existence of several
mechanisms that generate unphysical solutions. Among them, the
more important ones are:

 \indent - {\em  Negative masses:} in gravitational theories,
the solutions to field equations depend on a set of constants
that are related to the  masses of the system, but the (Newton
or Einstein) gravitational field theories do not contain neither
algorithms  nor constraint insuring that all the corresponding
masses of their exterior solutions are positive. For example, in
Newton gravitational theory, the number of unphysical solutions
with a finite number N of singularities (corresponding to a
distribution of N masses such that at least one of them is
negative) associated to every physical solution is given by $\,
\sum_{i=1}^N (\begin{array}{c} N\\i\end{array})= 2^N - 1\, $
\cite{unphysFinite}.

\indent - {\em Duality invariance:}  Maxwell equations at the
exterior of sources are  invariant by duality rotations, i.e. by
transformations of the form \,$F' = \cos\phi\, F + \sin\phi *\!F
\,.$ They are such that the sum of arbitrarily duality rotated
physical fields is not in general a duality rotation of a
physical field. Duality rotations introduce magnetic monopole
charges but we do not know neither algorithms nor constraint
equations allowing to know if a given (local, exterior) solution
$F$ to the field equations is able or not to generate, by a
duality rotation, a physical field. Now, for fields admitting a
finite number of singularities, the number of unphysical
solutions that correspond to every physical one is $R^N$
\cite{unphysFinite}.

\indent - {\em Advanced-retarded symmetry:} the principle of
causality and the finite character of the velocity of
perturbations in relativistic field theories lead to consider
physical phenomena as generated by retarded interactions
(electromagnetism and relativistic gravitation). But Maxwell
equations (and in some sense Einstein ones) admit indistinctly
retarded and advanced solutions, and no general algorithms  or
constraint equations are known to distinguish them. Furthermore,
Maxwell theory being linear, arbitrarily weighted sum of
advanced and retarded solutions, $\,\lambda F_{\rm retarded} +
\mu F_{\rm advanced}\, ,$ is a new solution, and one has also
$\,R^N\,$ unphysical solutions corresponding to every physical
one admitting N singularities \cite{unphysFinite}.

    In fact, we may conclude that {\em there are not known
complete and strict classical pure  field  theories}. The only
known example  of a strict theory is the little Newtonian
gravitational theory of one point particle \cite{Point
particle}; in spite of its restricted character, this example is
heuristically very rich, and allows to have an inkling of what a
complete and strict pure field theory looks like and in what
situations it may be useful. It is very striking that, apart
from this particular example, no other tentative had taken place
since the creation of the concept of field.

\vspace{4mm}

    {\bf d)} The construction of a new field theory, with
or without the above characteristics and whatever be the
motivations (to avoid singularities of their solutions, to
include in field form the equations of motion, to take directly
into account terms of self-interaction, to include additional
pure radiation solutions, etc) needs the following two tasks:

\indent \indent * to represent the physical field
quantities by pertinent mathematical objects,

\indent \indent * to find convenient differential
equations for them.

Paradoxically, the first of these points has been systematically
ignored. For this reason, the following section presents
comments on some aspects related to it.

%
%
\vspace{2mm}

\section{Mathematical Representation of Physical Fields}
\vspace{2mm}

\ \indent {\bf a)} In classical physics, the points of our
physical space or space-time are (locally) identified with the
points of the mathematical spaces $\,R^3\,$ or $\,R^4\,$. For
this reason, the vector character of, say, the position vector
is only a matter of mathematical definition.

But, once this identification is made, the  {\em particular}
tensor character, at every one of these points, of {\em any}
physical quantity has to be theoretically {\em founded} and
experimentally {\em verified} {\cite{Newtonfuerza}.}

    Thus, the adequacy of a physical quantity
with its formal or mathematical representation involves:

\indent $\bullet\,\,$ the consideration of the offer of
mathematical objects: scalars, vectors, tensors,  spinors,  etc,

\indent $\bullet\,\,$ the good comprehension of their invariant
ingredients as well as of the structure involving these
invariants  and

\indent $\bullet\,\,$ the experimental confrontation necessary
to  guarantee that the correct choice of representation has been
made.

\vspace{4mm}

 \indent {\bf b)} Concerning the first of these points, it is
important not to forget, as it is frequent, that the notion of
'tensor field' is always attached to a group, although
frequently implicit. Thus, in Newtonian mechanics the {\em
acceleration} of a particle is a vector for the whole Galileo
group  $\, G \,$ of coordinate transformations between inertial
observers, its {\em velocity} is  a vector  for the restricted
group $\, I \,$, $\, I \subset G \, ,$ of coordinate
transformations leaving invariant (internal or adapted) a given
inertial observer,  and its {\em position} is only a vector for
the more restricted group $\,I_0 \,$, $\, I_0 \subset I \subset
G \, ,$ of coordinate transformations that leave the origin
$\,O\,$ unchanged. Although trivial, this example shows that,
when exploring new, enlarged situations, the invariance group of
(the mathematical representation of) a physical quantity has to
be analysed carefully. In Special Relativity, this has not been
the case, for example, in generalizing Maxwell equations from
inertial observers to accelerated ones.

The electric field $\,e\,$ and the magnetic field $\,h\,$
measured by an inertial observer are assumed
\cite{supposedlyvectors} to be vectors under the above mentioned
group $\, I \,$ of  coordinate transformations adapted to that
observer. But they are not vectors under the whole Poincar\'e
group $\, P \,$, of coordinate transformations between arbitrary
inertial observers. As it is well known, there does not exist
any function of the sole electromagnetic quantities $\,e\,$ and
$\,h\,$ that be a tensor under $\, P \,.$ In order to construct
a tensor under $\, P \,,$ two important features are needed.
The first one is the {\em addition}, to the two quantities
$\,e\,$ and $\,h\,$, of the (unique) kinematical quantity
characterizing the inertial observer: its unit velocity $\,u\,.$
The second one is the {\em substitution}  of the search of the
vector character of the two initial ingredients $\,e\,$ and
$\,h\,$ by that of  the anti-symmetric tensor character of a
sole function  $\,F\,$ of the three ingredients $\,e\,,$ $\,h\,$
and $\,u\,.$ Both features lead to the well known result $\,
F\equiv F(e,h,u) = u \wedge e - *(u \wedge h)\, \, \cite{localcharts}. $

Faced to such a denouement and in such a spirit, the extension
of the above electromagnetic quantity $\,F\,$ to a larger group
of accelerated observers (be it in Minkowski space-time or in
the curved ones of General Relativity) involves the following
two physical questions: is the electromagnetic field $\,F_a\,$
measured by an accelerated observer a function $\, F_a(e,h,u)\,$
of its kinematical quantity $\,u\,$ alone, or does it depend
also on its (now non vanishing) acceleration $\,a\,,$ $\,F_a =
F_a(e,h,u, a)\,$ ? does the electromagnetic field quantity
$\,F_a\,$ measured by an accelerated observer remain a second
order anti-symmetric tensor? \cite{ueh}.

    Neither of the principles of relativity, covariance
or minimal coupling, in their usual formulations, allow to give
a clear answer to these questions. In fact, we have no other
arguments that mathematical simplicity or physical dogmatism to
clearly eliminate contributions of the above two aspects on the
mathematical representation of the electromagnetic field
quantities on the space-time.

\vspace{4mm}

 \indent {\bf c)} But even the assumption that the electric
and magnetic fields measured by an inertial observer are vectors
under his adapted group $\, I \,$ has to be submitted to
experimental agreement.
The assertion that they are vectors means that, if we measure
the force $\,{\bf f}_\alpha\,$ needed to cancel them at
different angles $\,\alpha\,,$ we must obtain the cosines law
$\,{\bf f}_\alpha = {\bf f}_0\cos \alpha \,.$

    Up to what precision such a law is true for the
electric and/or the magnetic fields?

Observe that a law different from the cosines law, even very
slightly different, will oblige to represent these fields by
means of geometric objects drastically different from vectors
(although analytically related \cite{butanalyticallyrelated}).

    Experiments such as, for example, the measure of the
ratio inertial/gravitational mass of a body are undoubtedly
interesting; but those trying to measure the adequacy of the
laws associated to the vector or tensor character of the
fundamental electromagnetic fields (among others) would provoke
a similar interest. Unfortunately, this is not the case at
present.

%
%

\vspace{2mm}

\section{Intrinsic Elements of the Electromagnetic Field}
\vspace{2mm}

\ \indent {\bf a)} In spite of the above comments, we shall
suppose here, unless otherwise stated,  that an electromagnetic
field in the space-time is (locally) described by a two-form
$\,F\,$ (second order anti-symmetric covariant tensor field)
such that, if $\,u\,$ is the unit velocity of an arbitrary
observer, the electric and magnetic fields for him are given by
\begin{equation}
e = {\rm i}(u)F \,\,\, , \,\,\, h ={\rm i}(u)*F \,\,\, ,
\end{equation}
where $\,{\rm i}\,$ stands for the interior product and $\,*\,$
for the Hodge operator associated to the space-time metric $\,g
\,$. Then,  one has equivalently $\,F =  u \wedge e - *(u \wedge
h)\,\cite{localcharts},$ where $\,\wedge\,$ stands for the
exterior product.

For many technical uses, it is sufficient to work with this
two-form $\,F\,$. But, at every point of the space-time, $\,F\,$
is an element of the tensor algebra over the real four
dimensional vector space.  Consequently, $\,F\,$ {\em cannot be
but a subset of vectors and numbers at every point,} that is to
say, a set of vector fields and scalar functions on the
space-time. In the tensor formalisms, these ingredients are
called the {\em invariants} or {\em intrinsic elements} of
$\,F\,$ \cite{subsetofvectors}.
\vspace{4mm}

\indent {\bf b)} It is well known that the independent scalar
functions of   F are two, usually chosen as
\begin{equation}
\phi \equiv {\rm tr}\,F^2  \,\,\, , \,\,\, \psi \equiv
{\rm tr}\,F\!*\!\!F \,\,\, ,
\end{equation}
where $\,{\rm tr}\,$ is the trace operator, and one has the
relations $\,\phi = 2(h^2-e^2)\,,$  $\,\psi = -4(eh)\,,$ which
reveal their implication on the fields $\,e\,$ and $\,h\,$;
they allow fixing, in the plane determined by these fields, one
of them as a function of the other.

But what it seems not known is the implications of these scalars
on the energy variables.

Remember that the physical components with respect to an
observer $\,u\,$ of the Minkowski energy tensor $\,T\,$ of
$\,F\,,$
\begin{equation}
T = \frac{1}{2}(F^2 + (*F)^2)
\end{equation}
are the {\em energy density} $\,\rho\,,$ the {\em Poynting
vector} $\,s\,$ and the {\em stress  tensor} $\, \tau\,,$
respectively given by
\begin{equation}
\rho \equiv {\rm i}^2(u)T \,\,\, , \,\,\, s
\equiv \bot(u){\rm i}(u)T \,\,\,, \,\,\,
\tau \equiv \bot(u)T  \,\,\,,
\end{equation}
where  $\,|s|\,$ is the  energy  across the  space-like unit
volume element per  unit  of time, $\,\perp\,$ denoting the
projector orthogonal to $\,u\,.$

Note that $\,\rho\,$ and  $\,|s|\,$  are {\em
relative-to-the-observers} quantities (i.e. not invariant). A
simple but interesting result is that {\em the difference of
theirs squares is an invariant quantity} \cite{cigual1}:
\begin{equation}
\label{ro2-s2-chi2}
   \rho^2 - |s|^2 = \chi^2  \,\,\, ,
\end{equation}
where
\begin{equation}
 \chi^2 \equiv \frac{1}{2^4}(\phi^2 + \psi^2)    \,\,\, .
\end{equation}
 We see that all the observers for which the Poynting vector
vanishes see the {\em same} energy density $\,\rho\,$, that this
energy density is a {\em minimum} and that this minimum amounts
the {\em invariant} quantity $\,\chi\,.$ This is why one is
naturally lead to give the following definition
\cite{CollSolerAlgebraic}:
\vspace{2mm}

{\bf Definition:} {\em The invariant $\,\chi\,$ is called the
{\em proper energy density} of the electromagnetic field, and
the observers that see it as their energy density, for which the
Poynting vector vanishes,  are said {\em at rest} with respect
to the electromagnetic field.}
\vspace{2mm}

All other observer will see an energy density $\,\rho\,$
corresponding to the rest energy $\,\chi\,$ incremented by the
Poynting energy $\,|s|\,$ according to (\ref{ro2-s2-chi2})
\cite{Poynting-kinetic}.

The stress tensor $\,\tau\,$ is also a relative-to-the-observer
quantity, related to the Poynting vector and to the energy
density by the eigen-value equation:
\begin{equation}
{\rm i}(s)\tau = \rho\,s  \,\,\, .
\end{equation}
A consequence of the above relation is that, in spite of the
relative-to-the-observer character of all the elements of this
equation, the other two eigen-values of  $\,\tau\,$ are
invariants. They just amount, up to a sign, the proper energy
density: $\, \pm\chi\,$ \cite{CollSolerAlgebraic}.
\vspace{4mm}

\indent {\bf c)} Among all the electromagnetic fields $\,F\,,$
there exists a particularly important class, the {\em pure
radiation  electromagnetic  fields.} Usually they are physically
characterized as those for which no observer sees a vanishing
Poynting vector,  or alternatively as those such that the
electric and magnetic components are orthogonal and equimodular.
But, by its physical clarity I prefer the following one
\cite{CollSoler}.
\vspace{2mm}

{\bf Definition:} {\em An electromagnetic field $\,F\,$ is a
{\em pure radiation field} if its proper energy density
vanishes, $\,\chi = 0\,,$  or alternatively if the whole energy
density is radiated as Poynting energy, $\,\rho  = |s|\,.$}
\vspace{2mm}

    All these definitions are equivalent and still equivalent
to any of the following relations:
\begin{equation}
s = \rho\, n \,\,\, , \,\,\, \tau = \rho\, n\otimes
n \,\,\, , \,\,\, F = \ell \wedge p \,\,\, ,
\end{equation}
where $\,n\,$ denotes the unit vector in the direction of the
Poynting vector and  $\,\ell\,$ and $\,p\,,$ $\,{\rm i}(\ell)p =
0\,,$ define respectively the {\em principal direction} and the
{\em polarization} of $\,F\,.$

Two-forms of the form $\,F = \ell \wedge p\,$ are called {\em
null.} For this reason, pure radiation electromagnetic fields
are also called {\em null fields}.

It is important to note that for every observer $\,u\,$ there
always exist a vector $\,\ell_u\,$ in the principal direction of
the null field such that $\,F = \frac{1}{\rho}(\ell_u \wedge
e)\,.$ It follows $\,\ell_u = \rho\, u + s \,,$ which gives the
physical interpretation of the principal direction: it is the
null lift of the Poynting vector $\,s\,$ with respect to the
observer $\,u\,.$
\vspace{4mm}

\indent {\bf d)} The non null electromagnetic fields are called
{\em regular}, and represent (radiating or not) electromagnetic
fields with Coulombian  part. They at rest or proper energy density
$\,\chi\,$ never vanishes, $\,\chi \not= 0\,,$ and they are of
the form
\begin{equation}
\label{FalfabetaU}
F = \alpha\, \ell \wedge m + \beta *(\ell\wedge m) \,\,\, ,
\end{equation}
where $\,\ell\,$   and  $\,m\,,$ are normalized null vectors,
i.e.:
\begin{equation}
\ell.\ell = m.m = 0 \,\,\, , \,\,\, \ell.m = 1 \,\,\, ,
\end{equation}
defining the {\em  principal  directions} of $\,F\,$, and
$\,\alpha\,$  and  $\,\beta\,$  are simply related to $\,\phi\,$
and   $\,\psi\,:$
\begin{equation}
\phi = 2(\alpha^2 - \beta^2)\,\,\, , \,\,\, \psi = -
4\alpha\,\beta \,\,\, .
\end{equation}
In terms of them, the electromagnetic proper energy density
$\,\chi\,$ is given by
\begin{equation}
\chi = \frac{1}{2}(\alpha^2 + \beta^2) \,\,\, .
\end{equation}

We have seen that the principal direction $\,\ell\,$ of a null
field is given by the null lift of the Poynting vector.

What is now, in the regular case, the physical interpretation of
the principal directions defined by $\,\ell\,$ and $\,m\,$? What
is their relation with the Poynting vector?

Paradoxically, these questions seem to have been never asked. In
order to answer them, but also to control on $\,F\,$ specific
properties of $\,\ell\,$ and $\,m\,,$ it is worthwhile to  solve
{\em intrinsically, covariantly and explicitly,}  the
eigenvector problem related to these principal directions
\cite{GenCovDed}.

The answer is generated by the operator $\,C\,$
\cite{CollFerrando},  given  in the following proposition.
\vspace{2mm}

{\bf Proposition 1} (Coll-Ferrando) : {\em The principal
directions $\,\{\ell\,\},$ and $\,\{m\}\,$ of a two-form  F  are
given by the vectors
\begin{equation}
\label{C(x)}
\ell = C(x) \,\,\, , \,\,\, m = ^tC(x) \,\,\, ,
\end{equation}
where the operator $\,C\,$, called the {\em principal
concomitant} of $\,F\,,$ is given by
\begin{equation}
\label{C}
C \equiv \alpha\, F - \beta\, *\!\!F + T + \chi\, g \,\,\, ,
\end{equation}
$\,^tC\,$ is its transposed and  $\,x\,$   is an arbitrary
time-like direction.}
\vspace{2mm}

Now, taking $\,x = u\,$ in (\ref{C(x)})one can prove the
following results \cite{CollSoler}:
\vspace{2mm}

{\bf Proposition 2} (Coll-Soler):  {\em The invariant directions
$\,\{\ell\}\,$ and $\,\{m\}\,$ of a regular electromagnetic
field $\,F\,$ are given by the null shifts $\,\ell_u\,$ and
$\,m_u\,$ respectively of the vectors
\begin{equation}
\ell_{\scriptscriptstyle \!\! \perp} = s - r \,\,\, , \,\,\,
m_{\scriptscriptstyle \!\! \perp} = s + r \,\,\, ,
\end{equation}
where $\,s\,$ is the Poynting vector and $\,r\,$ is given by:}
\begin{equation}
r \equiv \alpha\,e + \beta\,h \,\,\, .
\end{equation}
\vspace{2mm}

{\bf Corollary}:  {\em The Poynting vector $\,s\,$  with respect
to an observer  u of an electromagnetic field  $\,F\,$ is half
the sum  of the  projection of the invariant principal vectors
$\,\ell_u\,$  and  $\,m_u\,$ of $\,F\,$:}
\begin{equation}
s = \frac{1}{2}[\ell_{\scriptscriptstyle \!\! \perp} +
m_{\scriptscriptstyle \!\! \perp}] \,\,\, .
\end{equation}
\vspace{2mm}

This corollary strongly suggests a privileged decomposition of
an electromagnetic field into two pure radiation components. In
fact, one can prove the following result:
\vspace{2mm}

{\bf Proposition 3} (Coll-Soler):  {\em For every observer
$\,u\,$, the energy tensor $\,T\,$ of any regular
electromagnetic field $\,F\,$ may be obtained univocally as the
composition of two pure radiation energy tensors  along  the
principal directions of the field,
\begin{equation}
\label{T-Tl-Tm}
T = \frac{1}{2\chi}(T_{\ell_u}\,T_{m_u} + T_{m_u}\,T_{\ell_u})
-\chi\,g
\end{equation}
where  $\,T_{\ell_u}\,$  and  $\,T_{m_u}\,$   are given by}
\begin{equation}
T_{\ell_u} = \frac{2}{(\rho + \chi)}\, \ell_u \otimes
\ell_u \,\,\, , \,\,\,
T_{m_u} = \frac{2}{(\rho + \chi)}\, m_u \otimes m_u \,\,\, .
\end{equation}
\vspace{2mm}

 This result allows to say that the principal directions of
a regular electromagnetic field are the null shifts of
the Poynting vectors corresponding to the two pure radiation
electromagnetic fields whose composition (\ref{T-Tl-Tm})
generates the field.
\vspace{4mm}

\indent {\bf e)} From a technical point of view,  Proposition 1
constitutes a very interesting tool. It allows to solve inverse
problems concerning the necessary and sufficient conditions to
be verified by a tensor in order to insure particular
differential properties of some of its eigen-spaces. Up to now,
the only known result on such problems corresponded to a very
simple situation \cite{Haantjes}. As an example, we give here
for the electromagnetic field the following one, related to the
permanence of a pure radiation field \cite{CollFerrando}:
\vspace{2mm}

{\bf Proposition 4} (Coll-Ferrando):  {\em   An electromagnetic
field $\,F\,$ has a {\em geodesic} principal direction if, and
only if, its principal concomitant C, given by} (\ref{C}), {\em
satisfies
\begin{equation}
\label{geo}
{\rm tr}\{C\wedge {\rm i}(C)\nabla C\} = 0 \,\,\,  .
\end{equation}
where $\,C\,$ is considered as a vector valued one-form.}
\vspace{2mm}

In the last Section, similar techniques have been used to find
the non-linear generalization of Maxwell equations.
\vspace{4mm}

\indent {\bf f)} The real vector space structure of the
solutions of Maxwell equations means physically:

    - that Maxwell equations admit a {\em charge-scaling
law} for every electromagnetic field, and that this law is
multiplicative, and

    - that Maxwell equations admit a {\em superposition
law} for every two electromagnetic fields, and that this law is
additive.

The construction of a new electromagnetic field theory involves,
sooner or later, to ask about the existence of these two laws,
as well as about their respective multiplicative and additive
character.

For ``not-everything" theories, as would be the case of those we
are talking about here, the existence of such laws may be
epistemically required. It is thus their non-linear character
that has to be specified. Let us write them respectively in the
form
\begin{equation}
\label{leyescomposicion}
    \mu\bullet F = {\rm f}(\mu, F) \,\,\, , \,\,\,
    F\oplus G = {\rm g}(F,G)\,\,\,.
\end{equation}
Whatever be the tensor character of the field, the above
equations may be submitted to restrictions coming from desired
or suspected conditions. For example, the parameter $\,\mu\,$ in
(\ref{leyescomposicion}) may be restricted to take discrete
values corresponding to multiples of a elementary charge; the
function  $\,{\rm f}\,$ be such that $\,{\rm f}(-1, F) = -F\,$
for any $\,F\,$; or the function $\,{\rm g}\,$ be symmetric in
its arguments etc. But here we are interested only in the
general implications imposed by the two-form character of the
electromagnetic field $\,F\,.$ The corresponding general
expressions are given by the following proposition
\cite{compositionlaws}.

\vspace{2mm}

{\bf Proposition 5} (Coll-Ferrando):  {\em i) The more general
charge-scaling law for an electromagnetic two-form $\,F\,$ is of
the form
\begin{equation}
\label{lambdabullet}
\mu\bullet F = {\rm m}_F\,F + {\rm m}_{*F}\,*\!F
\end{equation}
where $\,{\rm m}_F\,$ and $\,{\rm m}_{*F}\,$ are functions of
the parameter  $\,\mu\,$ and of the two invariant scalars of
$\,F\,$.  ii) The more general superposition law for two
electromagnetic two-forms $\,F\,$ and  $\,G\,$ is of the form
\begin{eqnarray}
\label{foplusg}
     F\oplus G\, & = & \,\, {\rm p}_{\!F}\,\, F  + \,\,
     {\rm q}_G\,\,G \,\,   +  {\rm r}_{[F,G]}\,\,[F,G] \\
                 & + & \, {\rm p}_{*\!F}*\!\!F  + \,
     {\rm q}_{*G}\!*\!G \, +   {\rm r}_{*[F,G]}*\![F,G] \nonumber
\end{eqnarray}
where  $\,{\rm p}_{\!F}\,,$...$,\,{\rm r}_{*[F,G]}\,$ are
functions of the two invariant scalars of $\,F\,$, of the
corresponding scalars of $\,G\,$, and of the two mixed scalars
$\,{\rm tr\,FG}\,$
  and $\,{\rm tr\,F\!*\!G}\,,$ and $\,[F,G]\,$ is the
commutator of $\,F\,$ and $\,G\,,$ $\,[F,G]\equiv FG - GF\,.$}
\vspace{2mm}

The proof of expression (\ref{lambdabullet}) follows from the
fact that $\,\{F,*F\}\,$ is a basis, in the module of the
two-forms over the functions, for the powers of $\,F\,$, and
that of expression (\ref{foplusg}) follows from the fact that
the six two-forms of its second member form a basis for the Lie
algebra generated by $\,F\,$ and $\,G\,$ \cite{FernandoBaseSix}.

%
%

\vspace{2mm}

\section{Electromagnetic Field as a Lorentz Tensor}
\vspace{2mm}

\ \indent {\bf a)}    Some generic situations seem to indicate
that the linearity of the electromagnetic field equations is due
to the weakness of these fields in our ordinary experimental
conditions.

    Usually, those that take seriously this idea, try to
apply more or less reasonable criteria to find non linear
equations for the electromagnetic two-form $\,F\,$. This is to
forget that, if a weak electromagnetic field is represented by
the two-form $\,F\,,$ strong electromagnetic fields may change
not only its intensity, but even its tensor character.

Formally speaking, the first examples of such situations are
given by Group Theory: (the linear space of) the algebra of a
group is nothing but the set of ``weak elements" of the group.

In this sense, the electromagnetic case is very suggestive:
Maxwell equations, which are at the basis of Special Relativity,
structure the electromagnetic fields at every point as
anti-symmetric tensors, just like the elements of the algebra of
the Lorentz group that generates Special Relativity.

Thus, at least from a formal point of view, one is lead to test
the idea that the ``good" strong electromagnetic fields ought to be
represented by Lorentz tensor fields $\,L\,$, which, in the case
of little intensity, would reduce to simple two-forms $\,F\,$.

    The following two paragraphs do not pretend to present the
first ingredients of a new electromagnetic theory (which
nevertheless is in progress) but only to illustrate by a formal
example the general idea that {\em strong fields may change the
tensor character of a weak field representation,} and to show
that this idea is workable and interesting in some of its
consequences.
\vspace{4mm}

\indent {\bf b)} Thus, let us accept that strong electromagnetic
fields are correctly represented by Lorentz tensor fields
\cite{adimensional}.  Then,  ``near" the Maxwellian weak
electromagnetic solutions (technically: in the exponential
domain of the Lorentz Group), Lorentz tensors $\,L\,$
representing  generic  electromagnetic fields  are related
exponentially to their weak counterpart $\,F\,$ : $\,L = \exp
F\,.$

 The exponential and the logarithmic branches allow to exchange
the variables $\,F\,$ and $\,L\,$ in the exponential domain, and
also the corresponding equations. Nevertheless, the solutions to
these differential equations in terms of Lorentz tensors will
belong, not only to the exponential domain, or to the rest of
the connected-to-the-identity component but even to whole
disconnected group. The Lorentz tensors
corresponding to these last regions are not the exponential of
two-forms, so that, even imposing to $\,L\,$ the transformed
Maxwell equations, we will have much more electromagnetic
solutions than the exponential of the Maxwell ones.

But the new solutions corresponding to the disconnected
components will never reduce continuously, when varying the
integration constants, to the identity, i.e. will never become
vanishing electromagnetic fields. The possibility of the
physical existence of such fields is by itself very interesting
(see paragraph {{\bf 2.b}).}

 Let us already note that, for not strong electromagnetic
fields $\,L\,,$ for which two-forms $\,F\,$ exist such that $\,L
= \exp F\,,$ one has $\,L = g\,$ for vanishing $\, F\,.$ But the
metric  $\,g\,$ is nothing but the {\em inertia or gravitational
tensor field} of Minkowski or of general Riemann space-times
respectively: {\em Lorentz tensor fields are objects that may
simultaneously describe electromagnetic and gravitational
fields.}

For the above reasons, I believe that such fields are worthy of
deeper analysis.
\vspace{4mm}

\indent {\bf c)} The basic elements for such an analysis are the
functions \,{\em exp}\, and \,{\em ln}\,.  To handle them in
the context of a field theory, it is imperative to sum their
usual infinite series expansion.

Paradoxically, the sum of these series is still an open problem
even for many simple groups. For the sum of the exponential
series for the Lorentz group one has \cite{CollSanjoseexplog}:
\vspace{2mm}

{\bf Proposition 6}  (Coll-Sanjos\'e): {\em The exponential of a
two-form  $\,F\,$ in a space-time of metric tensor $\,g\,$ is
given by
\begin{equation}
\exp F = \epsilon_g\,g + \epsilon_T\,T+ \epsilon_{\!F} \,F +
\epsilon_{*\!F} \,*\!\!F \,\,\, ,
\end{equation}
where}
\begin{equation}
\begin{array}{c}
\vspace{1mm}
\epsilon_g = \frac{\displaystyle 1}{\displaystyle 2}(\cosh
\alpha + \cos \beta)\\
\vspace{1mm}
 \epsilon_T = \frac{\displaystyle 1}{\displaystyle
 \alpha^2+\beta^2}(\cosh \alpha - \cos \beta)\\
\vspace{1mm}
\epsilon_F = \frac{\displaystyle 1}{\displaystyle
\alpha^2+\beta^2}(\alpha\sinh \alpha +\beta \sin \beta)\\
\epsilon_{*F} = \frac{\displaystyle 1}{\displaystyle
\alpha^2+\beta^2}(-\beta\sinh \alpha +\alpha \sin \beta)
\end{array}
\end{equation}
\vspace{2mm}

 For the sum of the logarithmic series we present here,
for simplicity, only the result for non symmetric proper Lorentz
tensors on the principal branch \cite{othernranchs}.
\vspace{2mm}

{\bf Proposition 7}  (Coll-Sanjos\'e): {\em The logarithm of a
non symmetric proper Lorentz tensor field $\,L\,$ is given by
\begin{equation}
\ln L = {\rm h}_{^a\!L}\,\, ^a\!L +
\epsilon \,{\rm h}_{*^a\!L}\, *\!^a\!L
\end{equation}
where $\,^a\!L\,$ is the anti-symmetric part of $\,L\,$, $\,
\epsilon\,$ is the sign of the scalar $\,{\rm
tr}\,^a\!L*^a\!\!L\,$, and  $\,{\rm h}_{^a\!L}\,$ and $\,{\rm
h}_{*^a\!L}\,$ are the functions
\begin{equation}
    \begin{array}{r}
 {\rm h}_{^a\!L}\, = \frac{\displaystyle 1}{\displaystyle
 {\mu^2 - \nu^2}}\{(\mu^2-1)^{1/2}\arg\cosh \mu +
 (1-\nu^2)^{1/2}\arccos \nu\}\\
{\rm h}_{*^a\!L}\, = \frac{\displaystyle 1}
{\displaystyle{\mu^2 - \nu^2}}\{ (1-\nu^2)^{1/2}
\arg\cosh \mu - (\mu^2-1)^{1/2}\arccos \nu\}
    \end{array}
\end{equation}
of the invariant scalars $\,\mu\,$ and $\,\nu\,$ of $\,L\,$
given by}
\begin{equation}
    \begin{array}{r}
  \mu \equiv \frac{1}{4}\{{\rm tr}L + \sqrt{2{\rm tr}L^2
  - {\rm tr}^2L + 8}\} \,\,\,\,\,  \\\vspace{-3mm}
\\
  \nu \equiv \frac{1}{4}\{{\rm tr}L - \sqrt{2{\rm tr}L^2
  - {\rm tr}^2L + 8}\} \,\,\,.
    \end{array}
\end{equation}
\vspace{2mm}

In this scheme, some simple charge-scaling and superposition
laws on proper tensors homologous to the linear ones on their
associated two-forms may be imposed. The more natural ones are,
of course, those of the Lorentz group structure:
\begin{equation}
 \lambda\bullet L = \exp \{\lambda\,\ln L\}\,\,\, , \,\,\,
 L\oplus M = L\times M \,\,\, ,
\end{equation}
where $\,\times\,$ denotes the cross product
\cite{crossproduct}. The expression of this last law in terms of
the associated two-forms is known as the BCH-formula
(Baker-Campbell-Hausdorff), and its explicit and covariant
summation for the Lorentz group has been recently given. This
summation is, as it is due, of the form  (\ref{foplusg}) were
the particular values of the scalar coefficients $\,{\rm
p}_{\!F}\,$,...,$\,{\rm r}_{*[F,G]}\,,$ may be found in
\cite{Coll-SanjoseBCH}.

%
%
\vspace{2mm}

\section{Quadratic Electromagnetic Field Equations}
\vspace{2mm}

\ \indent {\bf a)} This Section presents a non-linear pure
field theory of electromagnetism, very likely the nearest to
Maxwell theory. As this last one, our non-linear theory supposes
that the electromagnetic field is a two-form $\,F\,$, but our
field equations for it turn out to be quadratic.

I consider this theory as slightly better than Maxwell one,
because it contains exactly the additional solutions we wanted
to have. But apart from this fact, it inherits all the other bad
aspects of Maxwell theory, particularly the duality invariance
and the advanced-retarded symmetry. In spite of that, I believe
it is worthwhile to present it:  as a slightly improvement on
Maxwell theory, of course, but also as an illustration of some
of the concepts and ingredients above mentioned and, overall,
because of its striking properties.
\vspace{4mm}

\indent {\bf b)} At the exterior of sources,  Maxwell equations
for an electromagnetic  two-form $\,F\,$ are
\begin{equation}
\label{Maxwell}
 dF = 0 \,\,\, , \,\,\, \delta F = 0 \,\,\, ,
\end{equation}
where $\,d\,$ and $\,\delta\,$ denote respectively the exterior
differential and the divergence (up to sign) operators.

For regular fields, expression (\ref{FalfabetaU}) may also be
written
\begin{equation}
 F = \alpha\,U + \beta\,*\!U
\end{equation}
where $\,U\,$  is a  unit  two-form,
\begin{equation}
 {\rm tr}\,U^2 = 2 \,\,\, , \,\,\, {\rm tr}\,U\!*\!U = 0 \,\,\, ,
\end{equation}
representing the {\em induced metric volume} on the time-like two-plane
$\,U\,$ of vectors $\,x\,$ such that  $\,{\rm i}(x)*U = 0\,.$
Maxwell equations admit for the weights $\,\alpha\,$ and
$\,\beta\,$ the {\em conditional system} in $\,U\,$
\cite{CondSystemFayos}

\begin{equation}
\label{MaxwellU}
    \begin{array}{c}\vspace{1mm}
 \delta\,[\delta \,U\wedge U - \delta *\!U\wedge *U]= 0  \,\,\, ,\\
 \delta\,[\delta\, U\wedge *U + \delta *\!U\wedge U]= 0    \,\,\, ,
    \end{array}
\end{equation}
\vspace{2mm}

\noindent from which   $\,\alpha\,$    and $\,\beta\,$  are
determined up to a constant related to initial values. Time-like
two-planes $\,U\,$ verifying (\ref{MaxwellU}) are called {\em
Maxwellian,} because they generate {\em all} the regular
solutions to Maxwell equations. Equivalently, Maxwell equations
for regular electromagnetic fields may be written, modulo
initial conditions, in the Rainich energy form \cite{Rainich}
\begin{equation}
\label{MaxwellT}
    \left. \begin{array}{lcr}\vspace{1mm}
{\rm tr}T = 0 & \,\,\,\,\,\, , \,\,\, \,\,\, & T^2\wedge
g = 0 \,\,\, \,\,\,\\
\delta T = 0 & , & d \,\frac{{\displaystyle *(T\times\nabla
T)}}{{\displaystyle
{\rm tr}T^2}}= 0 \,\,\, \,\,\,
    \end{array} \right \}
\end{equation}

\vspace{4mm}

\indent {\bf c)} It is important to note that neither
equations (\ref{MaxwellU}),
nor equations (\ref{MaxwellT}) are valid when $\,F\,$ is null.
For a null or pure radiation field,  $\,F = \ell\wedge p\,,$ the
eight  Maxwell  equations (\ref{Maxwell}) group in  {\em four}
sets of  {\em two}  equations implying  and  only  implying  the
following properties: \vspace{-2mm}
\begin{itemize}
    \item   the  null  direction is  geodesic,
\vspace{-2mm}
    \item  the  polarization  is  parallel  transported
along  the null direction,
\vspace{-2mm}
    \item the  null  direction is  distortion-free,
\vspace{-2mm}
    \item the  gradient of the energy density  is specifically
related to the  polarization vector.
\end{itemize}
\vspace{4mm}

\indent {\bf d)} This set of coupled equations appear as
excessively restrictive in some frequent situations.

It is the case in theoretical studies on wave-guides, where some
authors \cite{Harmuth} claim that Maxwell equations have an
insufficient number of pure  radiation solutions.

Also, in Special Relativity, meanwhile spherically symmetric
pure radiation electromagnetic fields are forbidden for evident
topological reasons, Maxwell equations are the sole responsible
of the banning of cylindrically symmetric pure radiation fields,
among others.

Even worse, there exists no vacuum gravitational space-times
in which Maxwell
equations admit generic pure radiation solutions; in other
words, a torch cannot bring light in General Relativity. This is
due to the Bel-Goldberg-Sachs theorem \cite{BelGolbergSachs},
that reduces drastically the existence of shear-free geodesic
null directions in curved space-times.
\vspace{4mm}

\indent {\bf e)} Pure electromagnetic fields, in particular
plane waves, like inertial observers, free particles or free
fields, are paradigmatic concepts in physics. Their existence
can only be proved locally and with rough precision, but their
importance resides both, in the simple concepts that they
involve and in the non-trivial constructions that they are able
to provide.

    On the other hand, the extraordinary success of Maxwell
equations in Physics is so enormous, that any answer allowing
them to save their form or their structure is in general
preferred. Thus, there exist some theoretical answers to the
above anomalies, still making Maxwell equations unchanged. But
we shall suppose here what is {\em the more direct} conclusion
from the above three points, namely that:

\vspace{3mm}

\centerline{ {\em   Maxwell equations contain insufficient
     pure radiation electromagnetic fields.}}
\vspace{4mm}

    Of the above four pairs of Maxwell equations for the null
case, the third pair is the responsible for the non existence of
cylindrical waves in Minkowski and generic waves in curved
space-times, and the fourth restricts severely the number of
solutions to those imposing a particular relation between the
orientation of the polarization and the intensity of the field.
There are these restrictions those that prevent to identify
electromagnetic pure radiation fields with beams of
electromagnetic rays. This is why we propose to substitute
Maxwell equations $\,{\bf M}(F)\,$  by a new set of equations
$\,{\bf S}(F)\,$ restricted  to the following {\em schedule of
conditions}:

\begin{description}
    \item{{\bf i.}} the pure radiation electromagnetic field
solutions of the new equations $\,{\bf S}(F)\,$ must be {\em
all} those having:
        \begin{itemize}
            \item their principal direction geodetic,
            \item their polarization parallel propagated,
        \end{itemize}
    \item{{\bf ii.}} the regular electromagnetic field
solutions to the new equations $\,{\bf S}(F)\,$ must differ as
little as possible from the corresponding  regular solutions to
Maxwell equations $\,{\bf M}(F)\,.$
\end{description}

\vspace{4mm}

\indent {\bf f)} There exists a natural class of operators on
algebras, called derivations. A derivation $\,{\bf d}\,$ of an
algebra $\,(+,\diamond )\,$ is an operator that verifies the
Leibniz rule for the product, $\,{\bf d}(f\diamond g) = {\bf
d}f\diamond g + \epsilon_f\,f \diamond {\bf d}g \,,$ where
$\,\epsilon_f\,$ is the parity sign of the element $\,f\,. $
Leibniz rule allows to associate to any other operator $\,{\bf
d\!\!\!\backslash}\,$  on the algebra an internal binary
composition law  $\,\{f,g\}\,,$ that we call the {\em Leibniz
bracket} of $\,{\bf d\!\!\!\backslash}\,$ with respect to the
algebra $\,(+,\diamond )\,,$
 by means of the relation $\,\{f,g\} + {\bf d\!\!\!\backslash}
 (f\diamond g) = {\bf d\!\!\!\backslash}f\diamond g +
 \epsilon_f\,f \diamond {\bf d\!\!\!\backslash}g \,.$ So, one
can say that, for a given algebra, an operator is a derivation
iff its Leibniz bracket vanishes.

 It is well known that the exterior derivative $\,d\,$ is a
derivation of the exterior algebra $\,(+,\wedge)\,,$ but that
the divergence operator $\,\delta\,$ is not.
\vspace{2mm}

 {\bf Definition:} {\em The} \,Schouten bracket
$\,\{F,G\}\,$ {\em of two exterior forms $\,F\,$ and  $\,G\,$ is
the Leibniz bracket of the divergence operator $\,\delta\,$ with
respect to the exterior algebra:
\begin{equation}
\{F,G\} \equiv \delta F \wedge G + (-1)^p\,F \wedge \delta\,
G - \delta \,(F\wedge G)
\end{equation}
where $\,p\,$ is the parity of $\,F\,. $}
\vspace{2mm}

With this instrument, and applying concepts of the preceding
Sections to write down these specifications in terms of the
electromagnetic field itself, one can proof:
\vspace{2mm}

 {\bf Proposition 8}  (Coll-Ferrando): {\em The equations
$\,{\bf S}(F)\,$ on the electromagnetic field two-form $\,F\,$
that satisfy the above schedule of conditions are:
\begin{equation}
\label{our system}
{\bf S}(F) \equiv \left\{
    \begin{array}{c}
\,\,\,\,\,
  \delta\,[F^2 + (*F)^2] = 0 \\
\\ \,\,\,\,\,
\{F,F \} + \{*F,*F\} = 0  \,\,\,\, ,
    \end{array}
            \right.
\end{equation}
where  $\,\{\,,\,\}\,$ is the Schouten bracket.}
\vspace{4mm}

\indent {\bf g)} Our new electromagnetic field equations {\bf
S}(F) verify the schedule of conditions in the {\em strongest}
sense: all the null two-forms with geodetic principal direction
and parallel propagated polarization are solutions of them, and
their regular solutions are {\em exactly} the regular Maxwellian
ones, in spite of the apparent difference between our equations
(\ref{our system}) and the Maxwell ones (\ref{Maxwell}).
Denoting by $\,\Sigma({\bf S})\,$ and $\,\Sigma({\bf M})\,$
respectively the space of solutions  of our system and that of
Maxwell equations, and by $\,\{ F_N \}\,$ the set of null
two-forms with geodetic principal direction and parallel
propagated polarization, one has
\begin{equation}
\Sigma({\bf S}) =  \Sigma({\bf M}) \cup \{ F_N \}
\end{equation}
Thus our new equations strictly make nothing but to add to
Maxwell equations the up to now missing pure radiation
solutions. But it is important to note that, if the
charge-scaling law remains the usual product by a number, now
the superposition law with ingredients in $\,\{ F_N \}\,$ is no
longer additive. In spite of our result (\ref{foplusg}) of
Proposition 5, for the moment we have been unable  to found it.

Maxwell succeed in formulating the action at a distance laws of
Coulomb, Biot-Savart, Ampere and Faraday in terms of the
electromagnetic force fields (adding its displacement current),
obtaining his celebrate equations \cite{MaxwellLinear}. Imagine
for a moment he tried to formulate them, in terms of the
energetic fields (energy density, Poynting vector, stress
tensor), instead of in terms of the force fields. Then, he would
obtained, instead of his equations, the Rainich ones
(\ref{MaxwellT}), valid only for regular fields; in other words,
{\em he would not discovered the electromagnetic character of
the light!}

The last of Rainich equations (\ref{MaxwellT}) is indeterminate
for  null energy  tensors, but, with the help of  null
two-forms, and expressed in term of them, this  indeterminacy
may  be {\em solved}, the result being  {\em our  equations}
(\ref{our system}).  In other words, {\em Maxwell could
discovered our equations!}

 Our equations are the necessary and sufficient conditions for
null principal directions to be geodetic with a parallel
transported polarization. These are the basic ingredients of the
geometric optics approximation. But Maxwell equations are
nothing but the {\em same} equations for {\em regular} fields,
so that we have:
\begin{quote}
      {\em the exact equations of the geometric optics
approximation of Maxwell  equations for null fields, are the
exact Maxwell equations for regular fields.}
\end{quote}

This result may be considered as a classical version of
Feynman's point of view on quantum electrodynamics.
\vspace{2mm}

\section{Conclusion}
\vspace{2mm}

    A little number of the many unclear aspects enveloping
electromagnetism have been commented. Some of them concern, more
generally, the very notion of (pure) field theory and the form
and properties of its equations (Section 2). And others affect
the double aspect, theoretical and experimental, of the tensor
character and texture of the electromagnetic field itself
(Sections 3 and 4). Its invariant elements, proper energy
density, observers at rest, space-like principal directions,
which paradoxically remained until now mere mathematical
variables, have been physically analysed (Section 4).

    It has been shown that strong fields could not only increase
the intensity of the ordinary ones but also change their tensor
character; and the example of Lorentz tensors, which, although
formal, shows the main features of this eventuality, has been
presented (Section 5).

    Finally, a physical application of some of the techniques
presented for the study of the texture of the electromagnetic
field has been given: an electromagnetic theory that generalizes
that of Maxwell, providing ``more light" than that contained in
it, but rigorously respecting all the not purely radiation ones
(Section 6). The new solutions could make our theory (more)
complete, if they were detected by appropriate experiments;
otherwise, lacking in physical meaning, they would make it less
strict. But in any case, the new theory suffers, at least, of
the same level of non-strictness than Maxwell theory.

    The problem of finding a complete and strict field theory
of electromagnetism remains therefore open. It has not been
possible to develop here other equally important unclear aspects
of the current field theories, but they constitute so many
reasons of dissatisfaction raised by the current notion of ``field
theory".

    The history of science reveals not only the brilliant
evolution of some ideas, but also and abundantly, the bad or
null evolutions of many others. It is the task of the physicist,
theorist or experimental, to correct this situation. I hope that
the simple results presented here be an incentive in this
direction.


\vskip 30pt

\begin{thebibliography}{99}

\bibitem{Collaborations} Many of the results and ideas included
here are the reflection of a friendly and fruitful collaboration
for a very long time with L. Bel, J.J. Ferrando, J.A. Morales,
F. San Jos\'e and A. Tarantola.

\bibitem{texture} We are here interested by the electromagnetic
fields in vacuum, irrespective of the sources that produce them.
The sources will be located at, or around, the singularities of
these vacuum fields, and an important but unsolved problem of
{\em all} the usual field theories is to find an algorithm
allowing to locate, from the knowledge of the {\em regular,
exterior field on a local, finite,  region}, its singularities
and the values of the charges associated to them. To our
knowledge, at present such an algorithm is only known for the
very simple gravitational theory of {\em one} Newtonian point particle
(see reference \cite{Point particle}).


\bibitem{Hehl} The other extreme point of view considers straightaway
electromagnetism in presence of matter, involving fields,
inductions, charge densities, currents and constitutive
equations. One of the best descriptions of this scenario, based in a
particularly elegant axiomatics, may be found in F. W. Hehl and Y.
N. Obukhov, {\em Foundations of Classical Electrodynamics}, in
press.

\bibitem{commentmatter}We work here with this {\em pure field}
notion better than with a matter model because in this last case
the equations obtained, on one hand, contain the vacuum ones
and, on the other, are ``closed" by the constitutive equations.
This amounts to say that they are equivalent to the vacuum ones
plus a particular class of electromagnetic matter models, so
that the ``texture part" of them is equally concerned by our
purposes.

\bibitem{UniversalLaw} B. Coll, {\em A Universal Law of
Gravitational Deformation for General Relativity,} in Proc. of
the ERE-98 Spanish Relativity Meeting in honour of the 65th
Birthday of Lluis Bel ``Gravitation and Relativity in General"
ed. J. Martin et al., World Scientific (1999). See also
http://coll.cc\,.

\bibitem{disconnected} The possibility to detach from particles
non-scaping parts of the field, may be related to the existence,
for Lorentz tensors, of four disconnected components. Only the
one containing the unit Lorentz tensor is directly related to
two-forms, that is to say, to ordinary electromagnetic fields.


\bibitem{NonLiMaxEqs} B. Coll and J.J. Ferrando, {\em Non Linear
Maxwell Equations} in Proc. of the ERE-93 Spanish Relativity
Meeting ``Relativity in General" \'Editions Fronti\'eres,
(1995).

\bibitem{verynull} ``Very null" in used here not as a mathematical
concept, but as an intuitive one, that wishes to express, as the
arguments that follow the text show, that solutions describing
physical phenomena constitute a set much lesser than the
``natural" subsets of null measure in the corresponding spaces.

\bibitem{unphysFinite} That means that, in the space of solutions,
for every point belonging to  $\,{\bf F}\,$, i.e. representing a
physical field, there are $\, 2^N - 1\, $ points in the
gravitational case and $\,R^N\,$ in the electromagnetic one,
that belong to $\,{\bf \widehat S}\,,$  i.e. that are unphysical
solutions.


\bibitem{Point particle} B. Coll and J.J. Ferrando, {\em The
Newtonian Point Particle,} in Proc. of the ERE-97 Spanish
Relativity Meeting ``Analytical and Numerical Approaches to
Relativity Sources of Gravitational Radiation" ed. by C. Bona et
al., Univ. de les Illes Balears, Spain (1998). See also
http://coll.cc\,.

\bibitem{Newtonfuerza} Already in the first two corollaries of
his Principia, Newton argues for the law of the parallelogram
for the composition of forces, and asserts that this law is
abundantly confirmed by mechanics. I do not know any experiment
conceived specifically to measure this concordance of
parallelograms and forces.

\bibitem{supposedlyvectors} Induced by an epistemic analysis of
vacuum Maxwell equations and the equivalence principle for
inertial observers. See the following paragraph c).

\bibitem{butanalyticallyrelated} They are the specific
formalisms for linear spaces that oblige to a drastic change
when passing from the cosine law to any other one. But it is
easy to see that in a general space of functions of a given
direction, the cosine law is dense in analytically related ones.

\bibitem{localcharts} In local charts $e_\alpha = u^\rho
F_{\rho\alpha}$, $h_\alpha =\frac{1}{2} u^\rho \eta_
{\rho\alpha\sigma\tau}F^{\sigma\tau}$, $\eta_
{\alpha\beta\gamma\delta}$ being the volume element of the
metric. Reciprocally, $F_{\alpha\beta} = u_\alpha e_\beta -
u_\beta e_\alpha - \eta_{\alpha\beta\rho\sigma}u^\rho h^\sigma
.$

\bibitem{ueh} It is to be noted that, although we call $\,F\,$
the {\em electromagnetic field}, this is {\em in fact} a short
cut for {\em electro-magnetic-observer field}: as the
expression $\, F(e,h,u) \,$ of the text reveals, one cannot
extract from $\,F\,$ the physical electromagnetic information
unless we fix the observer $\,u\,.$ In fact, one can show that,
in the generic case, the equation $\, F = F(e,h,u)\,$ for the
three variables  $\,u\,,$ $\,e\,,$ $\,h\,,$ admits, for every
arbitrary choice of one of them, a unique solution in the other
two.

\bibitem{subsetofvectors} That any tensor over a vector
space on a field K is nothing but a graph of collections of
vectors {\em weighted} by elements of K, is logically trivial
and algorithmically unexplored, up for second order tensors of
any type and third order tensors of the class of the structure
constants of a group. But both, for mathematical as well as
physical applications, it would be very interesting to develop
algorithms and canonical forms (extensions of the Jordan form)
for tensors of any order and type.

\bibitem{cigual1} We use everywhere c=1. In arbitrary
units $\,[\rho] =  [|s|][c]^{-1}\,.$

\bibitem{CollSolerAlgebraic} B. Coll and D. Soler,
{\em Algebraic properties of the electromagnetic field},
unpublished work.

\bibitem{Poynting-kinetic} Observe that the proper
energy density and the Poynting energy contribute to
the total energy in a form that, although quadratic,
is similar to that of the proper mass and kinetic energy
in the mechanical case. The difference is that, meanwhile for
mechanical systems there exists only {\em one} observer at rest,
for an electromagnetic field there exists a {\em one-parameter}
family.

\bibitem{CollSoler} B. Coll and D. Soler, {\em The radiation
parts of an electromagnetic field}, unpublished work.

\bibitem{GenCovDed} Intrinsic (i.e. expressed in terms of the
sole quantities present in the statement), covariant (i.e.
coordinate free) and  explicit (i.e. allowing to be checked by
direct substitution of the variables in the proposed
expressions) solutions to geometric or physical problems, or
{\em ICE solutions},  are very infrequent, in spite of their
great conceptual interest and practical advantages. The present
situation is an example. Surprisingly enough, in spite of the
old origin of linear analysis and the intense use of matrix and
tensor calculus in many branches of science, the general
solution to the eigenvector problem for matrices has been
considered but recently (note that we speak about the {\em
general solution}, i.e. the general expression, depending on the
matrix, of the invariant spaces corresponding to every
eigenvalue); see C. Bona, B. Coll and J. A. Morales, J. Math.
Phys., {\bf 33} (2) p. 670 (1992), for the four-dimensional
symmetric case, \cite{CollFerrando} for the corresponding
anti-symmetric case, and G. Sobczyk, The College Mathematical
Journal, {\bf 28} p. 27 (1997), for the general n-dimensional
case.

\bibitem{CollFerrando}If an electromagnetic field is a pure
radiation field at an instant (say, a plane wave), do Maxwell
equations insure that it will remain a pure electromagnetic
field in subsequent times? In contrast with a common opinion,
the answer is negative. See B. Coll and J. J. Ferrando, Gen.
Rel. and Grav., {\bf 20} (1) p. 51 (1988). In this paper, the
general eigenvector problem for two-forms in four-dimensional
space-times is solved.

\bibitem{Haantjes} J. Haantjes, Proc. Konink. Nederl. Akad.
Van Wetens. Amsterdam, {\bf A 58}(2), p. 158 (1955). He solved
the problem of obtaining the necessary ans sufficient
conditions, for symmetric tensors of type I, insuring that {\em
all} the eigen-vectors are integrable. What about the cases in
which {\em not all} them are integrable, or the cases in which
the tensor is not of type I, or those where the differential
system considered is not the integrability one? All these
questions remained open until now.


\bibitem{compositionlaws} B. Coll and J. J. Ferrando, {\em
Superposition Laws for the Electromagnetic Field}, unpublished
work.

\bibitem{FernandoBaseSix} B. Coll and F. San Jos\'e, J. Math.
Phys., {\em 37}, p. 5792 (1996); see also {\em Relative
positions of a pair of planes and algebras generated by two
2-forms in relativity}, in {\em Recent Developments in
Gravitation}, World Scientific (1991).

\bibitem{adimensional} In the present context it is supposed
that the electromagnetic fields are able to be described by
adimensional quantities. The ways and physical meanings that
allow this situation are theory-dependent, and will not be
discussed here.


\bibitem{CollSanjoseexplog} B. Coll and F. San Jos\'e, Gen.
Relativity and Grav., {\bf 22} p. 811 (1990).

\bibitem{othernranchs} For any proper Lorentz tensor and
any branch, see    \cite{CollSanjoseexplog}

\bibitem{crossproduct} Contraction over the adjacent spaces
of their tensor product, i.e. the matrix product of their
corresponding components.

\bibitem{Coll-SanjoseBCH} B. Coll and F.
San Jos\'e, Gen. Relativity and Grav.,{\bf 34} (9), p. 1345
(2002).

\bibitem{CondSystemFayos} B. Coll, F. Fayos and
J. J. Ferrando, J. Math. Phys., {\bf 28} (5), p. 1075 (1987).

\bibitem{Rainich} G. Y. Rainich, Trans. Am. Math. Soc.,
{\bf 27} p.106 (1925).

\bibitem{Harmuth} For a review, see H.F. Harmuth, {\em Propagation
of Non-sinusoidal Electromagnetic Waves} and also {\em Radiation
of Non-sinusoidal Electromagnetic Waves,}  Academic Press, 1986
and 1990 respectively)

\bibitem{BelGolbergSachs} L. Bel, Cahiers de Physique, {\em 16}
p.59-80 (1962); J.N. Goldberg and R.K. Sachs, Acta Phys. Polon.,
Suppl. {\em 22} p. 13 (1962).

\bibitem{MaxwellLinear} Observe that, if he tried to be rigorous
in this correspondence, he would be obliged to joint to its
linear equations the empirical quadratic inequalities $\,(e^2
-h^2)^2 + (e.h)^2 \not= 0\,$, verified in all the situations
implied by the above mentioned four laws. This empirical rigour
not only would break the linearity of his equations but would
forbidden the existence of electromagnetic waves!

\end{thebibliography}
\end{document}